\documentclass[%
 reprint,
showpacs,
floatfix,
amsmath,amssymb,
aps,
pra,
]{revtex4-1}

\usepackage{phd}
\usepackage{graphicx}% Include figure files
\usepackage{dcolumn}% Align table columns on decimal point
\usepackage{bm}% bold math
\begin{document}

\preprint{APS/123-QED}

\title{Photonic Quantum Operations via the Quantum Carburettor Effect}

\author{Jennifer C. J. \surname{Radtke}}%
 \email{jennifer.radtke@strath.ac.uk}
\author{Daniel K. L. \surname{Oi}}%
\author{John \surname{Jeffers}}
\affiliation{%
 SUPA Department of Physics, University of Strathclyde, Glasgow G4 0NG, UK
}%

\date{\today}

\begin{abstract}
The bosonic nature of light leads to counter-intuitive bunching effects. We describe an experimentally testable effect in which a single photon is induced through a highly reflecting beamsplitter by a large amplitude coherent state, with probability $1/e$ in the limit of large coherent state amplitude. We use this effect to construct a viable implementation of the bare raising operator on coherent states via conditional measurement, which succeeds with high probability and fidelity even in the high amplitude limit. 

\end{abstract}


\maketitle
\section{Introduction}

Operations on quantum states of light have a wide range of applications in quantum information processing and communication~\cite{Braunstein2005}, as well as being of fundamental interest~\cite{Kraus1983}. However, the operations that are easily implemented are limited to the class of Gaussian or linear operations~\cite{Braunstein2005}. Conditional evolution overcomes this limitation and provides a richer set of operations useful for manipulating continuous variable systems~\cite{Ban1996,Filip2005,Vitelli2010,Dakna1998}. Here, we show how to mimic the application of the bare raising operator to coherent state inputs using a beamsplitter, a single photon source, and a detector. In doing so we exploit a process that we call the quantum carburettor effect, whereby a strong coherent beam entrains the passage of a single photon (from an independent source) through a highly reflective beamsplitter with high probability, thereby elegantly highlighting the bosonic nature of light. This effect could also be used as a method to the characterize highly reflective beamsplitters and high mean photon number states. 

% photon subtraction\cite{Wenger2004} and addition\cite{Walker1986} (used eg in \cite{Parigi2007}), state comparison\cite{Andersson2006}, quantum scissors \cite{Pegg1998}, cat state generation \cite{Dakna1997,Ourjoumtsev2006}.
%Grangier \cite{Ferreyrol2010}, marek and filip \cite{Marek2010}, leuchs\cite{Usuga2010} and \cite{Eleftheriadou2013}
%^ Just so I can remember which one is which!

Conditional measurement-based evolution is a useful tool in discrete variable systems, such as the scheme by Knill, Laflamme and Milburn for efficient linear optical computing \cite{Knill2001}. Discrete systems have been extensively studied \cite{Nielsen}, but have limitations particularly apparent in communications, where loss may be significant. Continuous variable schemes show greater promise here \cite{Braunstein2005}, but while Gaussian states such as coherent or squeezed states are relatively well understood, non-Gaussian states and operations have been less well-studied \cite{Adesso2014}. Non-Gaussian operations are required for tasks such as entanglement distillation and error correction, essential to the use of continuous variables in information processing protocols. In the continuous variable regime conditional evolution allows non-Gaussian operations \cite{Braunstein2005}. There has significant interest in this approach \cite{Wenger2004,Walker1986,Parigi2007,Andersson2006,Pegg1998,Dakna1997,Ourjoumtsev2006}, as the alternative of using nonlinear optics typically succeeds with low probability due to the weakness of nonlinear susceptibilities. Operations based on conditional measurement can be concatenated and this can allow operations which would not be possible deterministically, such as probabilistic state amplification \cite{Ralph2009,Ferreyrol2010,Marek2010,Xiang2010,Usuga2010,Eleftheriadou2013,Donaldson2015}. Here we look at a different way to increase the energy of a quantum state, by implementing the ladder or bare raising operator \cite{Susskind1964}.

%In optics the Fock space is the set of photon number states.

We first introduce the bare raising operator, then describe the basic setup for photon addition using a beamsplitter. In contrast to previous work~\cite{Clausen1999,Escher2005}, our method works on an easily-generated state with Poisson photon number statistics such as the coherent state. The reflectivity of the beamsplitter is chosen to shape the photon number basis amplitudes of the output state to best match a photon-shifted coherent state. Our approach is not limited to coherent states as it does not rely on coherence between the input and the single-photon ancilla, and it will also provide a reasonably close implementation of the required operator on any state with  similar support to a coherent state. We show that the probability of successful implementation of the operation remains high even for large coherent state amplitudes -- the beamsplitter reflectivity in this case is close to unity, but the photon is still transmitted with high probability -- a process that we dub the quantum carburettor effect. We examine the effect of imperfect detector efficiency and show that the operations and quantum carburettor effect persist for experimentally feasible values. We also consider a multi stage scheme.

\section{The bare raising operator}

\begin{figure}
\input{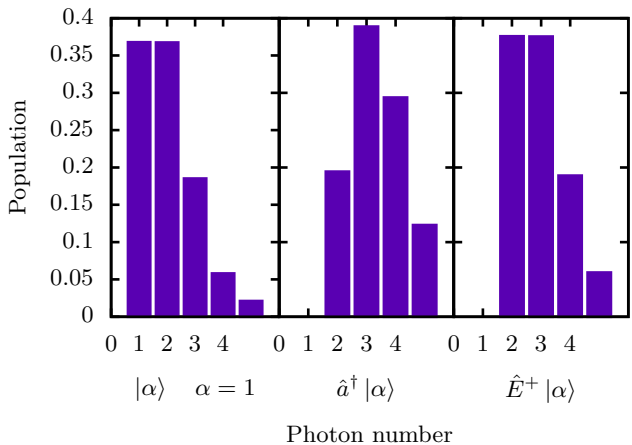}
\vspace{1ex}  
\caption{Comparison of the action of $\cre$ and $\add$. Figure shows Fock basis populations for a coherent state $\ket{\alpha}$, $\cre\ket{\alpha}$ (after normalization) and $\add\ket{\alpha}$  with $\alpha=1$. Note that $\cre$ increases the relative probability of higher excitation numbers (bosonic enhancement), whereas $\add$ preserves the relative populations.\label{pncompare}}
\end{figure}

The bare raising and lowering operators, sometimes known as the Susskind-Glogower operators~\cite{Susskind1964}, act on the space of harmonic oscillator energy eigenstates to shift the amplitudes of a state of a system up or down the ladder by exactly one quantum without modifying their relative amplitudes. They are
\begin{equation}
\hat{E}^+ = \sum_{n=0}^\infty |n+1\rangle \langle n|,\label{addexact}\qquad
\hat{E}^- = \sum_{n=1}^\infty |n-1\rangle \langle n|,
\end{equation}
with $\add=(\sub)^\dagger$. Note that while $\sub\add = \ident$, $\add \sub = \ident - |0 \rangle \langle 0|$, because acting with $\hat{E}^-$ on the ground state has no support. Unlike the more usual creation and annihilation operators $\cre = \sum_{n=0}^\infty \sqrt{n+1}|n+1\rangle \langle n|$ and $\ann= \sum_{n=1}^\infty \sqrt{n} |n-1\rangle \langle n|$ the bare operators do not introduce $\sqrt{n}$ bosonic enhancement factors, as shown in Fig.~\ref{pncompare}.  The corresponding operations are identical in their actions on a Fock state, but for a superposition or mixture of Fock states the difference between these two operations can be clearly seen. $\add$ only shifts the Fock basis amplitudes up to a higher photon number, whereas after normalization $\cre$ increases the amplitudes of larger Fock states relative to lower ones. Therefore the bare raising and lowering operators can be used to shift the Fock state amplitudes of a quantum state up or down whilst preserving coherence, with the obvious exception of the ground state information being lost when the state is lowered.

These operators have long been of theoretical interest \cite{Susskind1964}, as they can be used in applications such as generation and manipulation of nonclassical states~\cite{Steinhoff2014} or their characterization~\cite{Zou2006}. Also, the bare raising operator $\add$ is a Fock-space equivalent of the first Hilbert Hotel type operation \cite{Potocek2015}, which demonstrates the mathematical concept of infinity by an apparent paradox: a fully occupied hotel with infinite rooms can accommodate one more guest by moving everyone up by one room.

\begin{figure}
\begin{tikzpicture}[scale=0.9]
%Locate and draw the beamsplitters
\node (BS1) at (0,0) {};
\draw (BS1) -- +(45:0.75cm) node[anchor=south] {$t_1$} -- +(225:0.75cm) ;
\node (BS2) at (1.5,0) {};
\draw (BS2) -- +(45:0.75cm) node[anchor=south] {$t_2$} -- +(225:0.75cm);
\node (BS3) at (3,0) {};
\draw (BS3) -- +(45:0.75cm) node[anchor=south] {$t_3$} -- +(225:0.75cm);
\node (BSN) at (6,0) {};
\draw (BSN) -- +(45:0.75cm) node[anchor=south] {$t_{n-1}$} -- +(225:0.75cm);

%Locate and draw the detectors
\node (D1) [above=of BS1] {};
\draw ($(D1) - 0.5*(0.5,0)$) arc (180:0:0.25) -- cycle;
\draw ($(D1) + 0.5*(0,0.5)$) -- ($(D1) + (0,0.5)$) node[anchor=south] {0};
\node (D2) [above=of BS2] {};
\draw ($(D2) - 0.5*(0.5,0)$) arc (180:0:0.25) -- cycle;
\draw ($(D2) + 0.5*(0,0.5)$) -- ($(D2) + (0,0.5)$)node[anchor=south] {0};
\node (D3) [above=of BS3] {};
\draw ($(D3) - 0.5*(0.5,0)$) arc (180:0:0.25) -- cycle;
\draw ($(D3) + 0.5*(0,0.5)$) -- ($(D3) + (0,0.5)$)node[anchor=south] {0};
\node (DN) [above=of BSN] {};
\draw ($(DN) - 0.5*(0.5,0)$) arc (180:0:0.25) -- cycle;
\draw ($(DN) + 0.5*(0,0.5)$) -- ($(DN) + (0,0.5)$)node[anchor=south] {0};

%Horizontal modes
\draw[->] ($(BS1) - (1,0)$) node[anchor=east] {$\1$} -- (BS1);
\draw[->] (BS1) -- (BS2);
\draw[->] (BS2) -- (BS3);
\draw[->] (BS3) -- ($(BS3) + (1,0)$);
\draw[loosely dotted] ($(BS3) + (1.25,0)$) -- ($(BSN) - (1.25,0)$);
\draw[->] ($(BSN) - (1,0)$) -- (BSN);
\draw[->] (BSN) -- ($(BSN)+(1,0)$) node[anchor=west] {$\ket{N}$} ;

%Vertical modes
\draw[->] ($(BS1) - (0,1)$) node[anchor=north] {$\1$} -- (BS1);
\draw[->] ($(BS2) - (0,1)$) node[anchor=north] {$\1$} -- (BS2);
\draw[->] ($(BS3) - (0,1)$) node[anchor=north] {$\1$} -- (BS3);
\draw[->] ($(BSN) - (0,1)$) node[anchor=north] {$\1$} -- (BSN);
\draw[->] (BS1) -- (D1);
\draw[->] (BS2) -- (D2);
\draw[->] (BS3) -- (D3);
\draw[->] (BSN) -- (DN);

%Some more dots
\draw[loosely dotted] ($(D3) + (1.25,0.2)$) -- ($(DN) - (1.25,-0.2)$);
\draw[loosely dotted] ($(BS3) + (1.25,-1.2)$) -- ($(BSN) - (1.25,1.2)$);
\end{tikzpicture}
\caption{Setup of scheme in \cite{Escher2005} for the synthesis of large Fock states. The scheme succeeds when all detectors have no counts. The beamsplitters each have a different transmission probability to optimize the probability of a success, given success of the previous stage.\label{escfig}}
\end{figure}
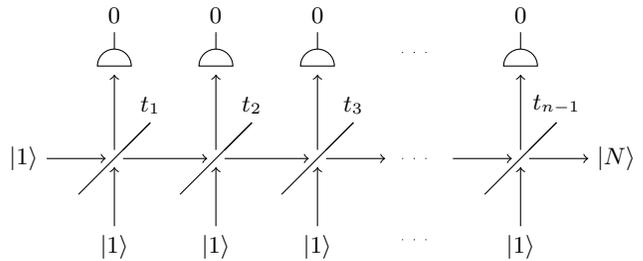

A scheme to synthesize arbitrary Fock states using beamsplitters and conditional measurement with single photon inputs was considered in 2005 by Escher \emph{et. al.} \cite{Escher2005} . Their system consisted of a cascade of beamsplitters, each combining a single photon Fock state with the output of the previous beamsplitter. This is depicted in Fig.~\ref{escfig}. The scheme requires $N$ single photons to make a Fock state $\ket{N}$, and succeeds when all $N-1$ perfectly efficient detectors do not fire. They found that the maximum probability of a given detector not firing to be
\begin{equation}
P(0) = \left(\frac{n}{n+1}\right)^n \label{esceq}
\end{equation}
for a Fock state $\ket{n}$ and single photon input, leading to a $\ket{n+1}$ output. This occurs with a beamsplitter of transmission coefficient 
\begin{equation}
t_n = \sqrt{\frac{n}{n+1}}  \text{ .}
\end{equation}

The standard creation operator $\cre$ can be implemented approximately, either using postselected spontaneous parametric down conversion \cite{Clausen2001}, or using a beamsplitter, single photon and detector. We use the latter approach here, but we aim to implement the bare raising operator $\add \ket{n} = \ket{n+1}$ instead. We do this by an appropriate choice of the reflection coefficient. Recently, experimental implementations of $\add$ using cavity QED~\cite{Oi2013} or circuit QED~\cite{Govia2012,Joo2016} have been proposed that may allow practical applications. Linear optical implementations of higher-order Hilbert-Hotel operations exist, but not in the Fock basis~\cite{Potocek2015}. Here we replace the complexity of these proposals with linear optics in the Fock basis and postselection.

\section{Implementation of $\add$ with a single beamsplitter}
\subsection{Setup}
%Basic system diagram
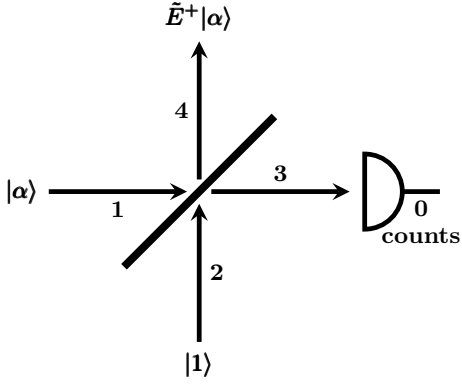
\begin{figure}[htbp]
\begin{tikzpicture}
\beamsplitter{$\pmb{\ket{\alpha}}$}{$\pmb{\1}$}{}{$\pmb{\addap{\ket{\alpha}}}$}
\detector{4.2}{2}{0.5}
\path (4.7,2) -- node[anchor=north,align=center] {\textbf{0}\\ \textbf{counts}} (5.2,2);
\end{tikzpicture}
\caption{\label{BS} Setup of scheme to implement $\add$. A coherent state in mode 1 and a single photon in mode 2 are the inputs. Measurement of no photocounts at a perfectly efficient detector in mode 3 gives an approximate implementation of $\add$ on the input coherent state, denoted $\addap\ket{\alpha}$ in mode 4. The beamsplitter has transmission coefficient $t$ and reflection coefficient $r$.}
\end{figure}

%\footnote{Note that $t$ and $r$ are reversed compared to reference \cite{Escher2005}.} - put this back if recompiling with bibtex
The essential setup used throughout is shown in Fig.~\ref{BS}, with coherent state and single photon inputs and measurement in one output mode \cite{Note1}. A coherent state of mean photon number $|\alpha|^2$ can be written as:
\begin{equation}
\ket{\alpha} = e^{\frac{-|\alpha|^2}{2}} \sum_{n=0}^\infty \frac{\alpha^n}{\sqrt{n}} \ket{n} \text{ ,}
\end{equation}
where $\{\ket{n}\}$ is the set of Fock states. The beamsplitter relations are
\begin{equation}
\begin{pmatrix}
\cre_1 \\ \cre_2
\end{pmatrix} = \begin{pmatrix}
|t| e^{\imag \phi_T} & - |r| e^{-\imag \phi_R} \\ |r| e^{\imag \phi_R} & |t| e^{-\imag\phi_T}
\end{pmatrix} \begin{pmatrix}
\cre_3 \\ \cre_4 \end{pmatrix} \text{ ,}\label{BSeq}
\end{equation}
where $|t|^2 + |r|^2 =1$. The operation implemented by this setup will be denoted $\addap$ to distinguish it from the ideal $\add$ operation. So that $\addap$ does not change the relative phase of the photon number states, we choose the convention $\phi_R = \pi$ and $\phi_T = 0$.

%General expression
The joint input state to the beamsplitter is 
\begin{equation}
\ket{\alpha_1}\ket{1_2} = e^{\frac{-|\alpha|^2}{2}} \sum_{n=0}^\infty \frac{\alpha^n}{\sqrt{n}} \ket{n}_1\ket{1}_2 \equiv \sum_{n=0}^\infty q_n \ket{n}_1\ket{1}_2 \text{ ,}
\end{equation}
where $q_n = e^{\frac{-|\alpha|^2}{2}} \frac{\alpha^n}{\sqrt{n!}}$. This can be written in terms of the input mode creation operators acting on the joint vacuum state:
\begin{equation}
\ket{\Psi_\text{in}} = \sum_{n=0}^\infty q_n \frac{\cre_1 \,^n}{\sqrt{n!}} \cre_2 \ket{0 0}_{12} \text{ .}
\end{equation}
Application of the beamsplitter transformation in Eq. \eqref{BSeq} gives the joint output as:
\begin{IEEEeqnarray}{rCl}
\ket{\Psi_\text{out}} &=& \sum_{n=0}^\infty \frac{q_n}{\sqrt{n!}} \left(|t|e^{\imag\phi_T} \cre_3 - |r|e^{-\imag \phi_R}  \cre_4\right)^n \nonumber\\
& & \times \left( |r| e^{\imag \phi_R} \cre_3 + |t| e^{-\imag \phi_T} \cre_4\right) \ket{0 0}_{34} \text{ ,}
\end{IEEEeqnarray}
which can be expanded to give:
\begin{eqnarray}
\hspace{-2ex}\ket{\Psi_\text{out}} & = & \sum_{n=0}^\infty  \sum_{k=0}^n  \frac{q_n \sqrt{n!} (-1)^k}{k! (n-k)!} \nonumber \\ & & \times \left[ A_1 \sqrt{(n-k+1)!k!} \ket{n-k+1}_3 \ket{k}_4 \right.\nonumber \\
& & \left. + A_2 \sqrt{(n-k)!(k+1)!} \ket{n-k}_3 \ket{k+1}_4 \right] \text{ ,}\label{cohout}
\end{eqnarray}
where
\begin{eqnarray}
A_1 = A_1(n,k) & = & |t|^{n-k}|r|^{k+1} e^{\imag(n-k)\phi_T -\imag (k-1)\phi_R} \\
A_2 = A_2(n,k) & = & |t|^{n-k+1}|r|^k e^{\imag(n-k-1)\phi_T -\imag k \phi_R} \text{ .}
\end{eqnarray}

After conditioning on no counts at the detector in mode 3 (\emph{i.e.} $k=n$, 2nd term only), we find the normalized output is
\begin{equation}
\addap \ket{\alpha} = \frac{1}{\sqrt{P(0)}}\sum_{n=0}^\infty q_n |t| |r|^n \sqrt{n+1} \ket{n+1}_4 \text{ ,} \label{addapout}
\end{equation}
where the success probability $P(0)$ is given by
\begin{equation}
P(0) \sum_{n=0}^\infty |q_n|^2 |t|^2 |r|^{2n} (n+1) \text{ .} \label{p0}
\end{equation}

The output state from the beamsplitter can be tuned by adjusting the value of $|r|$. We adjust $|r|$ to match the desired state $\add \ket{\alpha}$ more closely. We do this by optimizing $F_1 = |\bra{\alpha}(\add)^\dag \addap \ket{\alpha}|^2$, which is the fidelity of the beamsplitter-implemented operation with the exact operation. 

%This is not a phase dependent effect: the phase of the two inputs need not be precisely matched, only the frequency and polarization. The operation implemented is non-Gaussian and thus of interest in a range of applications, including sequential measurement schemes and entanglement distillation. 

%%Plot to show the limit of 1/e
%Revert to this version (combining the two other figures) if space becomes an issue.
%\begin{figure}[htb]
%\centering
%\input{nondetectoracc_new.tex}
%\caption{Graph of the success probability (green, solid) of the $\addap$ operations for coherent state $\ket{\alpha}$ input, with limit $1/e$ (grey, solid), and the optimal value of the reflection probability $|r|^2=\sin^2\theta$ (purple, dashed). The fidelities (red, dotted) are the fidelity of the beamsplitter implemented $\addap\ket{\alpha}$ compared with the perfect $\add\ket{\alpha}$ (upper) and of $\cre \ket{\alpha}$ compared with $\add\ket{\alpha}$ (lower), showing the distinguishability of the two operations.\label{cohstateopt}\todo[inline]{discuss distinguishability} }
%\end{figure}

%Analytical limits

\subsection{Results}
\subsubsection{Perfect Detectors}
\begin{figure}[htbp]
\centering
\input{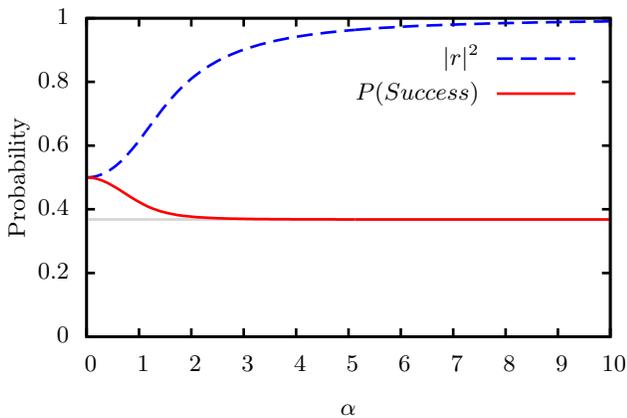}
\caption{Success and reflection probabilities for the $\addap$ operation. The probability of no counts (red, solid) for the beamsplitter-implemented $\addap$ operations for coherent state $\ket{\alpha}$ input is plotted, with limit $1/e$ (grey, solid), and the optimal value of the reflection probability $|r|^2$ (blue, dashed).\label{nondetgraph_prob}}
\end{figure}

The most striking result is the behavior of the success probability for the $\addap$ operation in the high $n$ or $\alpha$ limit. The limit of Eq. \eqref{esceq} as $n \rightarrow \infty$ is $P(0) = 1/e$ for high-$n$ Fock states. This result also holds for $\alpha \rightarrow \infty$, as shown numerically in Fig.~\ref{nondetgraph_prob}. Thus a single photon is, with high probability, transmitted through a highly-reflecting mirror by an intense light beam. We call this the quantum carburettor effect. Note that this is not due to any coherence between the photon and the coherent state, as would be the case in a Mach-Zender interferometer for instance.

The quantum carburettor effect can clearly be seen by the probability limit in Fig.~\ref{nondetgraph_prob}. With a beamsplitter of the optimum reflection probability $|r_\text{opt}|^2$, the probability for the detection of zero photons with a perfect detector tends towards the limit $1/e \approx 0.37$. The optimum reflection probability tends towards a perfectly reflecting mirror, but even in this limit a single photon can be transmitted with probability $1/e$ if the appropriate amplitude coherent state is also incident on the beamsplitter.

\begin{figure}[htbp]
\centering
\input{nondetectoracc_new_part2short.tex}
\caption{Fidelity of the $\addap$ operation compared to $\cre$. The fidelities of the normalized output state $\addap\ket{\alpha}$ (red, solid) and normalized $\cre \ket{\alpha}$ (blue, dashed) are shown, compared to the ideal state $\add\ket{\alpha}$. Fidelities are $|\bra{\alpha}(\add)^\dag \addap \ket{\alpha}|^2$ and $|\bra{\alpha}\ann \addap \ket{\alpha}|^2$ respectively (with $\cre$ states suitably normalized). The reflectivity of the beamsplitter was chosen to maximize the fidelity of the output state with $\add \ket{\alpha}$.\label{nondetgraph_fid}}
\end{figure}

We show the fidelity of the conditional output state after normalization with the ideal $\add \ket{\alpha}$ in Fig.~\ref{nondetgraph_fid}, and also for comparison the fidelity of the state $\cre\ket{\alpha}$ with the ideal $\add \ket{\alpha}$.  The different effect of $\add$ and $\cre$ is clearest around $\alpha\approx1$, and the conditional output state has a much higher fidelity with $\add\ket{\alpha}$ than the state $\cre\ket{\alpha}$.

Numerical results indicate that the reflection probability $|r_\text{opt}|^2$ for implementing the $\add$ operation with the highest fidelity coincides closely with that for the highest success probability for large values of $|\alpha|$ ($\alpha>\approx 2.5$).  
To calculate the reflection probability corresponding to the maximum success probability, we differentiate Eq. \eqref{p0} w.r.t. $|r|^2$ and set equal to zero.
% \begin{eqnarray}
% \frac{\ud P(0)}{\ud |r|^2} & = & e^{-|\alpha|^2} \sum_0^\infty \frac{|\alpha|^{2n}}{n!} (n+1) \left[ - |r|^{2n} + \right. \nonumber \\ & &  \qquad \qquad \left. (1-|r|^2) n |r|^{2(n-1)}\right] \nonumber\\
% 0 & = & e^{-|\alpha|^2} \left[ \sum_0^\infty -\frac{(|\alpha|^{2}|r|^2)^n}{n!} (n(n-1) + (3n+1)) \right. \nonumber\\ & & \qquad \qquad \left. + \sum_0^\infty \frac{|\alpha|^{2n}}{n!} |r|^{2(n-1)}(n(n-1) + 2n) \right] \nonumber\\
% & & = e^{-|\alpha|^2} e^{|\alpha|^2 |r|^2} \left[ -(|\alpha|^2|r|^2)^2 -3|\alpha|^2 |r|^2 -1 \right. \nonumber\\ & & \qquad \qquad \left. + |\alpha|^4 |r|^2 + 2 |\alpha|^2 \right] \text{ .}
% \end{eqnarray}
Taking the positive solution for $|r|^2$ leads to
\begin{equation}
|r|^2 = \frac{|\alpha|^2 - 3 + \sqrt{|\alpha|^4 + 2|\alpha|^2 +5}}{2|\alpha|^2} \label{cohoptprob} \text{ ,}
\end{equation}
which is valid above $|\alpha|^2 = 0.5$. Hence in the high $|\alpha|$ limit Eq. \eqref{cohoptprob} is a good approximation of the optimal reflection probability.

\subsubsection{Inefficient Detectors}
\begin{figure}[tbp]
\centering
\input{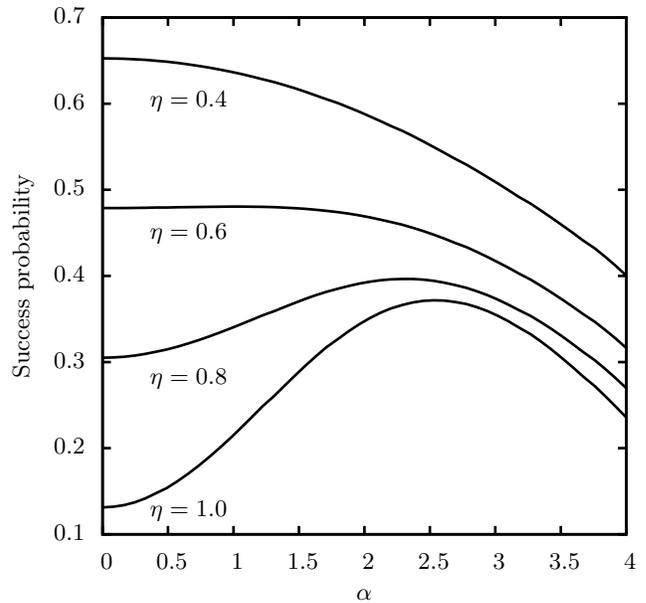}
\caption{Effect of detector efficiency on $\addap$ success probability. The probability of zero photocounts is plotted against $\alpha$ for fixed reflection probability $|r|^2 = 0.869$. Each line represents a different detector efficiency: from bottom to top, $\eta = 1.0,0.8,0.6,0.4$.\label{alphahumpgiveneta}}
\end{figure}

A major consideration for an experimental implementation would be the robustness to detector inefficiency. This can be accounted for by using the normally ordered measurement operator $:\exp(-\eta \cre \ann):$ in place of a simple projection \cite[6.10]{Loudon2000}, where $\eta$ is the detector efficiency (the probability that a photon incident on the detector will be counted). Numerical calculations shown in Fig.~\ref{alphahumpgiveneta} indicate that the model is not affected severely by the presence of moderate inefficiency. Numerical work was done in python with the aid of the QuTiP package \cite{qutip}.

\begin{figure}[!htbp]
\centering
\input{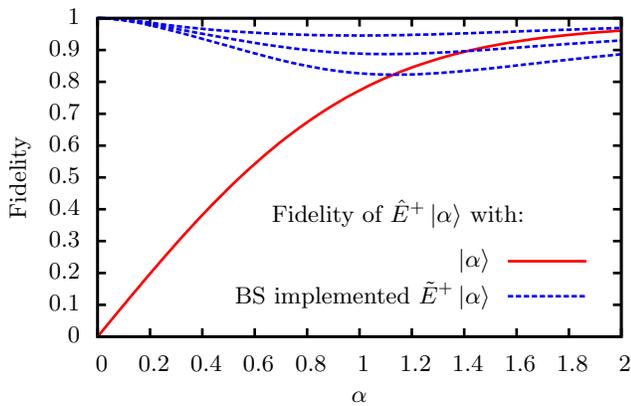}
\caption{Comparison of the $\addap$ scheme with doing nothing when detectors are inefficient. The fidelity of the original coherent state $\ket{\alpha}$ with its photon number raised state $\add\ket{\alpha}$ (red,solid) and the fidelity of the normalized beamsplitter operation with $\add\ket{\alpha}$ are shown for values of the detector inefficiency $\eta = 0.8,0.6,0.4$ from top to bottom (blue, dashed).\label{basefid}}
\end{figure}

%Remove optimising: phrase as an upper bound on $\alpha$ possible given detector efficiency. 
Whilst moderate detector inefficiency does not prevent implementation of $\add$, for $\eta < 1-1/e$ it introduces an upper bound on possible input coherent state amplitudes $\alpha$. This is due to the switch between two regimes: for low $\alpha$, the beamsplitter-implemented operation gives a better fidelity, while for higher $|\alpha|$ the loss of fidelity due to the inefficiency means that simply reflecting the input coherent state gives a better fidelity. In that case the success probability is the same as the detector inefficiency, although the operation cannot be said to be implemented. The boundary between these two regimes for various values of $\eta$ is depicted in Fig.~\ref{basefid} as the point at which the `do nothing' fidelity exceeds the achievable fidelity with the current set up.

\section{Cascaded Operation}
% Intro to 2BS approach
With a more elaborate approach, it may be possible to improve on the scheme in the previous section. Here we consider a straightforward extension of the implementation of $\add$, with extra components to correct a failed operation.

\subsection{Setup}
\begin{figure}
\begin{tikzpicture}
\beamsplitter{$\pmb{\ket{\alpha}}$}{$\pmb{\ket{1}}$}{}{}
\detector{4.2}{2}{0.5}
\node (D1) at (5.2,2) {};
\path (4.9,2) -- node[anchor=south] {\textbf{D1}} (5.2,2);
\node [anchor=north west,align=center] at (5.3,2) {\hspace{-3ex}\textbf{$\pmb{x=0, 1}$ or $\pmb{>1}$}\\ \hspace{-2ex}\textbf{count(s)}};
\node at (3.2,3.2) {\textbf{BS1}};
\draw[line width = 3pt] (1,6) -- node (centerii) {}  (3,8) node[anchor=south west] {\textbf{BS2}};
\draw[line width = 1.8pt, -stealth] (0,7) node[anchor=east] {$\pmb{\ket{1}}$} -- node[anchor=north] {6} (centerii);
\draw[line width = 1.8pt, -stealth] (2,4.5) node[anchor=north,draw] (delay) {Switch} -- node[anchor=west] {\textbf{5}} (centerii);
\draw[line width = 1.8pt, -stealth] (delay) -- node[anchor=south] {If $x>1$} ++(-2,0) node[anchor=east] {\textbf{Fail}};
\node[anchor=west] at ($(delay)+(0,0.7)$) {If $x=1$};
\draw[line width = 1.8pt, -stealth] (centerii) -- node[anchor=south] {\textbf{8}} (4,7) node[anchor=west] {$\pmb{\ket{\psi_2}}$};
\draw[line width = 1.8pt, -stealth] (centerii) -- node[anchor=east] {\textbf{7}} (2,9) node[anchor=south] {};
\vdetector{2}{9.2}{0.5}
\path (2,9.7) -- node[anchor=west] {\textbf{0 counts}} node[anchor=east] {\textbf{D2}} (2,10.2);
\draw[->, dotted, thick] (D1.center) .. controls ($(D1.west) + (2,0)$) and ($(delay.south east) + (1,0)$) .. node[anchor=west,align=right] {Feedforward \\ $x$} (delay.south east);
\draw[-stealth, line width = 1.8pt] (delay.east) -- node[anchor=south] {If $x=0$} (4,2 |- delay.east) node[anchor=west] {$\pmb{\ket{\psi_1}}$};
\end{tikzpicture}
\caption{Cascaded $\addap$ operation. The lower beamsplitter (BS1) is as in the previous section. If a failure occurs at BS1), heralded by the single photocount at detector D1, the scheme now continues. The failed state is input into the upper beamsplitter (BS2), with a second ancilla single photon, and a second attempt made to implement $\addap$, with result $\ket{\psi_2}$ in the event that no photocounts occur at detector D2.\label{2BSSchema}
}
\end{figure}
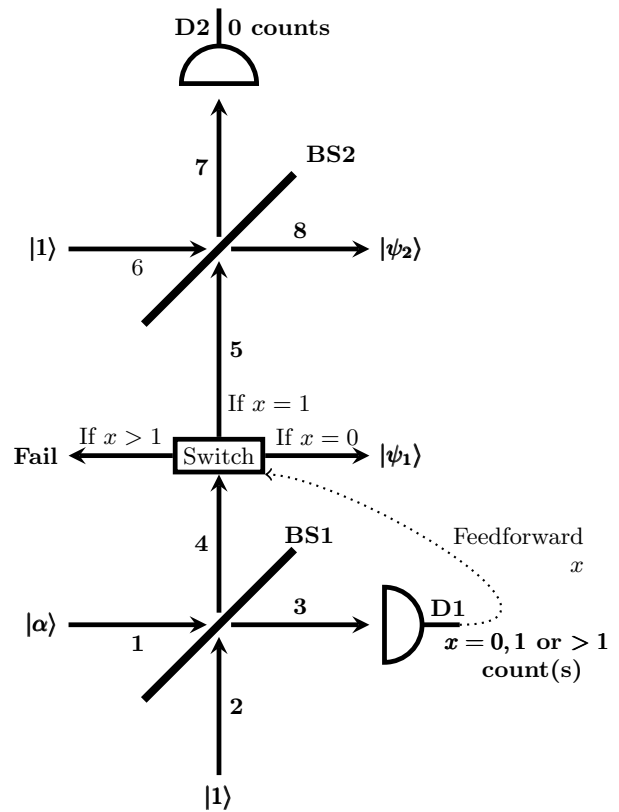

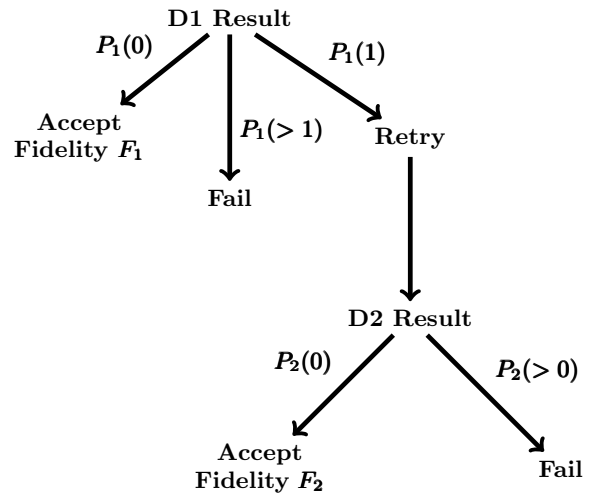
\begin{figure}
\begin{tikzpicture}[scale=2]
%Sort text/positions
\node (D1) at (1,0) {\textbf{D1 Result}};
\node (F1) at (0,-0.8) [align=center] {\textbf{Accept} \\ \textbf{Fidelity }$\pmb{F_1}$};
\node (fail1) at (1,-1.2) {\textbf{Fail}};
\node (retry) at (2.2,-0.8) {\textbf{Retry}};
\path (2.2,-2) node (D2) {\textbf{D2 Result}} +(-1,-1) node [align=center] (F2) {\textbf{Accept} \\ \textbf{Fidelity} $\pmb{F_2}$} +(1,-1) node (fail2) {\textbf{Fail}};
%Draw lines and add probabilities
\draw[line width = 1.8pt, ->] (D1) -- node[anchor=south east] {$\pmb{P_1(0)}$}(F1);
\draw[line width = 1.8pt, ->] (D1) -- node[anchor=north west] {$\pmb{P_1(>1)}$}(fail1);
\draw[line width = 1.8pt, ->] (D1) -- node[anchor=south west] {$\pmb{P_1(1)}$} (retry);
\draw[line width = 1.8pt, ->] (retry) -- (D2);
\draw[line width = 1.8pt, ->] (D2) --  node [anchor=south east] {$\pmb{P_2(0)}$} (F2);
\draw[line width = 1.8pt, ->] (D2) -- node[anchor=south west] {$\pmb{P_2(>0)}$} (fail2);
\end{tikzpicture}
\caption{Probability tree associated with the correction scheme in Fig.~\ref{2BSSchema}. D1 and D2 refer to the detectors as labelled in Fig.~\ref{2BSSchema}, while $P_j(0)$ indicates the no-count probability at the $j$th detector, and $P_j(1)$ the single-count probability. If more than 1 count occurs, we do not attempt correction.\label{tree}}
\end{figure}

The operation fails with 1 or more photocounts in mode 3. For the case of 1 count, we attempt the operation again on the failed output. This requires feedforward from the first detector, and an additional single photon, beamsplitter and detector as in Fig.~\ref{2BSSchema}. Figure \ref{tree} shows the possible measurement results at each detector, with their associated outcomes and probabilities. This may improve the success probability of the operation.

If there is no count at detector 1, then the first output state $\ket{\psi_1}$ is accepted as before. If there is a count, a correction is attempted. When there is no count at detector 2 the correction is accepted and the unnormalized output state $\ket{\psi_2}$ in mode 8 is
\begin{equation}
\sum_{n=0}^\infty q_n |r_1|^{n-1} (n |t_1|^2 - |r_1|^2) |t_2||r_2|^n \sqrt{n+1} \ket{n+1}_8 \text{ ,} \label{2BSstate}
\end{equation}
where subscript 1 refers to the first beamsplitter parameters and 2 to the second. The normalization gives the probability of an initial failure (1 count at detector 1) and then an accepted correction (no counts at detector 2). 

The fidelity measure used is the mean fidelity of accepted output states:
\begin{equation}
F_{mean} = \frac{P_1(0)F_1 + P_1(1) P_2(0) F_2}{P_1(0)+P_1(1) P_2(0)} \text{ ,}
\end{equation}
where $P_j(0)$ indicates zero counts at the $j$th detector and therefore an accepted output state at the relevant beamsplitter, $P_j(1)$ indicates one photocount with the possibility to correct the state, and $F_j$ is the fidelity of that state. The total success probability is simply $P_1(0)+P_1(1)P_2(0)$. 

\subsection{Results}

\begin{figure}[htbp]
\centering
\input{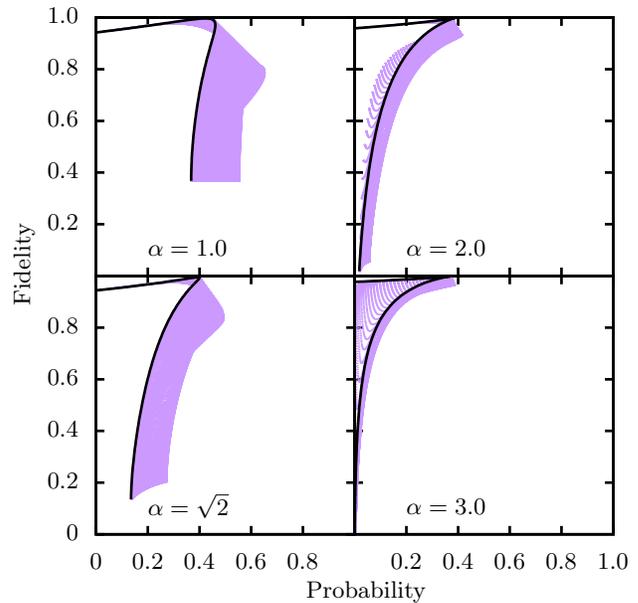}
\caption{The success probability-fidelity trade off. The mean fidelities of the accepted output states $\addap\ket{\alpha}$ with ideal state $\add\ket{\alpha}$ for $\alpha=1, 2,  \sqrt{2}$ and $3$ are shown, against the probability of obtaining that outcome in a 2 beamsplitter setup. Points (purple, light) cover the full range of possible pairs of beamsplitters and indicate the accessible region of fidelities and probabilities using this approach. The line (black, bold) indicates outcomes solely using a single.
\label{doubleBSgraph}}
\end{figure}

Figure \ref{doubleBSgraph} shows, for various input $\ket{\alpha}$, the possible mean fidelities against their probabilities. Points cover the full range of beamsplitter pairs, whilst the solid line indicates use of beamsplitter 1 alone. It can be seen that the best fidelity is obtained using a single beamsplitter. For lower $\alpha$ an increase in probability is possible, but this results in a large loss of fidelity. This indicates that the strategy delivers only marginal improvement at best. We explore the reason for this later.

% Results of 2BS approach

\section{Discussion and conclusions}
From the basic scheme it is clear that $\add$ is an experimentally-viable, high probability non-Gaussian operation with potential applications in continuous variable tasks such as entanglement distillation. The application of $\add$ makes coherent states nonclassical. This is clearly seen by noting that $\add \ket{\alpha}$ has vanishing amplitude in the zero photon component, a sign of nonclassicality \cite{Mandel1995}. 

In the presented implementation of $\add$, the maximum success probability and optimal fidelity are achievable near simultaneously; there is no need to compromise one to improve the other as is the case with the implementation of $\cre$. More elaborate optical schemes may improve the success probability, as there is in principle no limit to the coherent state amplitude on which $\add$ may be implemented with high probability. This was demonstrated theoretically in \cite{Oi2013} in a cavity QED setup.

%Experimental implementation (no diagram needed)
To observe the quantum carburettor effect experimentally, the main components required are a single photon source and laser suitable for interference, a beamsplitter and a photodetector. As the measurement is conditioned on zero photocounts, a non-photon-number-resolving detector should be sufficient. This would allow the investigation of the $1/e$ probability limit for the transmission of a single photon with matched beamsplitter $|r_\text{opt}|^2$ and $\alpha$. 

\begin{figure}[htbp]
\centering
\input{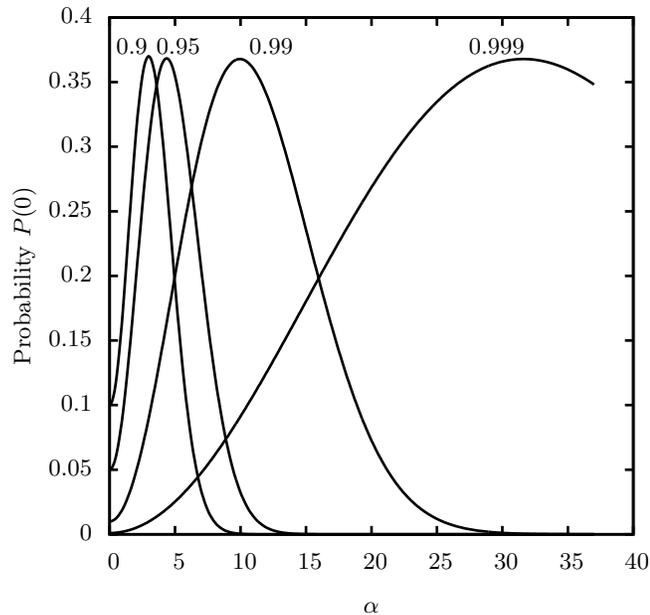}
\caption{Characterizing highly reflective beamsplitters using the quantum carburettor effect. The probability of zero photocounts against $\alpha$ is shown for a fixed reflection probability. Each line represents a beamsplitter corresponding (left to right) to reflection probabilities $|r|^2 = 0.9$, $0.95$, $0.99$ and $0.999$.  \label{characbs}}
\end{figure}

An alternative experimental application for the quantum carburettor effect is shown in Fig.~\ref{characbs}. The quantum carburettor effect and the existence of an optimal reflection probability dependent on $\alpha$ leads to a peak in probability just above $1/e$ when $\alpha$ is adjusted. This effect can be used to characterize highly reflecting beamsplitters, or alternatively with a variable beamsplitter to characterize high $\alpha$ coherent states.  

\begin{figure}
\input{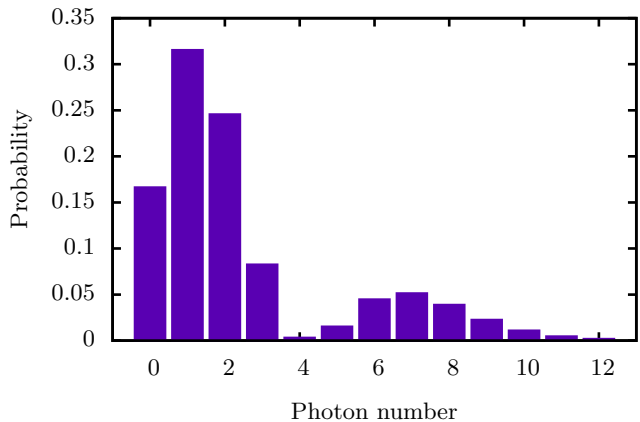}
\caption{Output state after a failed operation with 1 count in mode 3 and a perfect detector. The input state was a coherent state of amplitude $\alpha=2$.\label{pnfail}}
\end{figure}

The poor performance of attempts to correct the failed operation with a second beamsplitter can be linked to the two photon interference and the production of a `hole' in the photon number distribution of a coherent state, an effect described by Escher et al \cite{Escher2004} and depicted in Fig.~\ref{pnfail}. As $\alpha$ increases, the optimal beamsplitter reflection coefficient $|r_\text{opt}|$ for the $\add$ operation on a coherent state tends towards the reflection coefficient required to create a hole around $n=|\alpha|^2$ found by Escher et al. Figure \ref{pnfail} shows this effect for $\alpha=2$, with the optimal beamsplitter for the $\add$ operation. This lack of amplitude at $|\alpha|^2$ severely impacts on the fidelity of any attempted recovery; while any divergence from this $|r_\text{opt}|$ in the first beamsplitter reduces the probability and fidelity of an initially successful operation. Hence, the attempt at correcting a failed operation is not particularly successful. Future work could consider other schemes to improve the success probability for $\add$, or look at ways to implement other non-Gaussian operations. 

A scheme to implement the bare photon raising operator has been presented, using linear optics and conditioning on a measurement outcome. Through this the quantum carburettor effect (interference between a high amplitude single photon and coherent state) has been introduced and various applications considered, including characterizing beamsplitters and large coherent state amplitudes. A possible extension of the scheme was considered and found to give little improvement in success probability. These operators are experimentally realizable and of relevance in quantum information processing in continuous variable systems. 

\section*{Acknowledgments}
The authors acknowledge useful discussions with Brian Gerardot and Adetunmise Dada. J. C. J. Radtke acknowledges support from the EPSRC Doctoral Training Grant, University of Strathclyde.

\end{document}